\documentstyle[12pt]{article}
\begin{document}
\baselineskip 0.3in
\begin{center}
{\Large {\bf Pressure dependence of the low-frequency dielectric 
constant of KNbO$_3$}}

\vspace{0.4in}
D. Errandonea\footnote{e-mail address:daniel@ges1.fisapl.uv.es} 
and E. Moya

\vspace{0.2in}
Departamento de F\'{\i}sica Aplicada, Universidad de Valencia\\ 
C/ Dr.  Moliner 50, E-46100 Burjassot (Valencia), Spain
\end{center}

\vspace{1in}
\begin{abstract}
The effect of pressure on the low-frequency dielectric constant, 
$\epsilon_0$, of single crystals of KNbO$_3$ is investigated by 
means of capacitance measurements. The dielectric constant increases 
with pressure up to 22.5 kbar, where it exhibits a large value 
($\epsilon_0$ = 5000), and then decreases. This change in its 
behaviour is related to a phase transition induced by pressure. 
On decompression, the samples do not revert back to the ambient 
pressure phase.    
\end{abstract}

\vspace{0.5in}
\noindent
PACS numbers: 77.80.B, 77.84.D, 64.70.-p

\pagebreak

Potassium niobate (KNbO$_3$), which is a prototype of a ferroelectric 
material, at ambient conditions presents an orthorhombic structure. 
Recently, KNbO$_3$ has been studied under hydrostatic pressure 
showing that this compound undergoes some structural 
changes\cite{6,7}. By Raman scattering several phase transitions have 
been reported\cite{7}. The pressure-induced phases are different from 
those induced by temperature at ambient pressure. In particular, a 
displacive type phase transition has been proposed at 20 kbar. This 
transition is mainly characterized by softening of a transverse 
optical (TO) mode with wavelength $\omega_{TO}$ = 50 cm$^{-1}$.
However, its transition pressure, $P_t$, is not well established since 
the mode frequency cannot be detected above 20 kbar due 
to the disappearance of this mode into the continuous background.
The interest in more detailed information about this phase transition
stimulated the present measurements of the low-frequency dielectric, 
$\epsilon_0$, constant under pressure up to 32 kbar, at room
temperature.

The KNbO$_3$ crystal samples used in the experiments were grown from a 
melt comprised of a mixture of K$_2$CO$_3$ and NbO$_5$. 
High-resistivity single crystals were oriented along [010] direction,
being $\epsilon_0$ in this orientation larger than in the other axis. 
Samples for the capacitance measurements were cut and polished into 
slabs 100-200 $\mu$m thick and 2x2 mm$^2$ in size. Silver electrodes 
were vacuum evaporated on the large sample faces. Ohmic contacts were 
made by soldering silver wires to the electrodes with high-purity 
indium. The capacitance of the samples was measured by 
using  a high-accuracy ($>$ 0.1$\%$) capacitance meter and shielded 
leads. High-pressure was generated by a  Bridgmann cell which has been 
described in earlier studies\cite{8}. In this case, we have used 
tungsten carbide anvils, 27 mm in diameter, without steel binding 
rings and sodium chloride was the pressure-transmitting medium. The 
pyrophyllite gaskets employed, which were treated at 720 $^o$C during 
one hour in order to get suitable mechanical properties, were 0.5 
mm in thickness with a hole of 9 mm in diameter.  The pressure was 
determined by calibration of the load applied to the anvils  against 
known fixed points.

The capacitance, $C$, at a given pressure, $P$, is given by:
\begin{eqnarray}
C(P) = \epsilon_0(P) \frac{A(P)}{d(P)} \quad ,                                  
\end{eqnarray}

\noindent
where  $A$ is the area of the silver electrodes and $d$ is the slab 
thickness. Then, through Eq. (1) we can obtain $\epsilon_0(P)$ from 
$C(P)$, provided one takes into account the changes in the sample 
dimensions due to the compression. For this purpose, we have used a 
Murnaghan equation of state with the bulk modulus $B_0$=1420 kbar and 
$B'_0$=4 as deduced by x-ray diffraction\cite{10}. Figure 1 shows the 
results of $\epsilon_0$ as a function of the applied pressure for a
sample with $d$ = 100 $\mu$m. There it can be seen that, under 
compression, $\epsilon_0$ increases until it reaches its maximum 
value ($\epsilon_0$ = 5000) at about the pressure of 22.5  
kbar. In particular, $\epsilon_0$ increases nearly linear up to
around 15 kbar, above this pressure the slope $d \epsilon_0$/$d P$ 
changes from 75 kbar$^{-1}$ to 180 kbar$^{-1}$. Finally,
above 22.5 kbar $\epsilon_0$ decreases monotonically. We think, that 
the drastict change in the behaviour of the low-frequency dielectric 
constant at 22.5 kbar is originated by the phase transition induced by 
pressure reported in Ref. \cite{7}. The roundness of the observed peak 
may be related to the quasi-hydrostatic conditions at which we carried 
out the experiments. Note that unavoidable uniaxial stresses and 
sample deformations, due to the solid pressure-transmitting medium 
used, can contribute to reinforce crystal instabilities. Fig. 1 also 
shows that on decompression $\epsilon_0$  does not increase as 
expected for a reversible phase transition. This attests an apparently 
non-reversible phase transition, where the high-pressure phase 
remains stable at ambient conditions after releasing the pressure. 
The results have been tested for their reproducibility in five
different samples. Then, by averaging the pressure where the maximum 
of $\epsilon_0$ occurs we estimated that $P_t$ = 22.5 $\pm$ 0.5 kbar

The results obtained below $P_t$ can be checked
by calculating indirectly the Gr\"uneisen parameter $\gamma$
of the 50 cm$^{-1}$ TO mode. In ferroelectric materials, as a first
approximation, the relationship between $\epsilon_0$ and the
TO mode wavelength is given by the
Lyddane-Sachs-Teller relation. Under this assumption, it is easily
to show that the Gr\"uneisen parameter of $\omega_{TO}$ is given by: 
\begin{eqnarray}
\gamma=\frac{-B_0}{2}\left(\frac{d \ln \epsilon_0}{d P}\right)_{P=0} 
\quad.
\end{eqnarray}

\noindent
Then, from our results we get $\gamma$ = -29. This large value of
$\gamma$ indicates that a small change in volume pruduces a relatively
large change in $\omega_{TO}$ as was previously stablished\cite{7}.

In conclusion, our results confirm that a non-reversible
phase transition induced by pressure occurs at around 20 kbar. 
In addition, we have estimated that $P_t$ = 22.5 $\pm$ 0.5 kbar. 

\vspace{0.2in}
\noindent
{\large{\bf Acknowledgments}}

\noindent
{\small This work was supported through Spanish Government CICYT  
Grant No. MAT95- 0391. One of the authors (D:E:) wishes also to
thank Programa Mutis for its financial support.}

\vspace{0.3in}
\noindent
{\Large {\bf Figure captions}}

\vspace{0.125in}
{\bf Figure 1:} Pressure dependence of the static dielectric 
constant of KNbO$_3$. Symbols represent the data obtained during 
compression ($\circ$) and decompression ($\bullet$). 

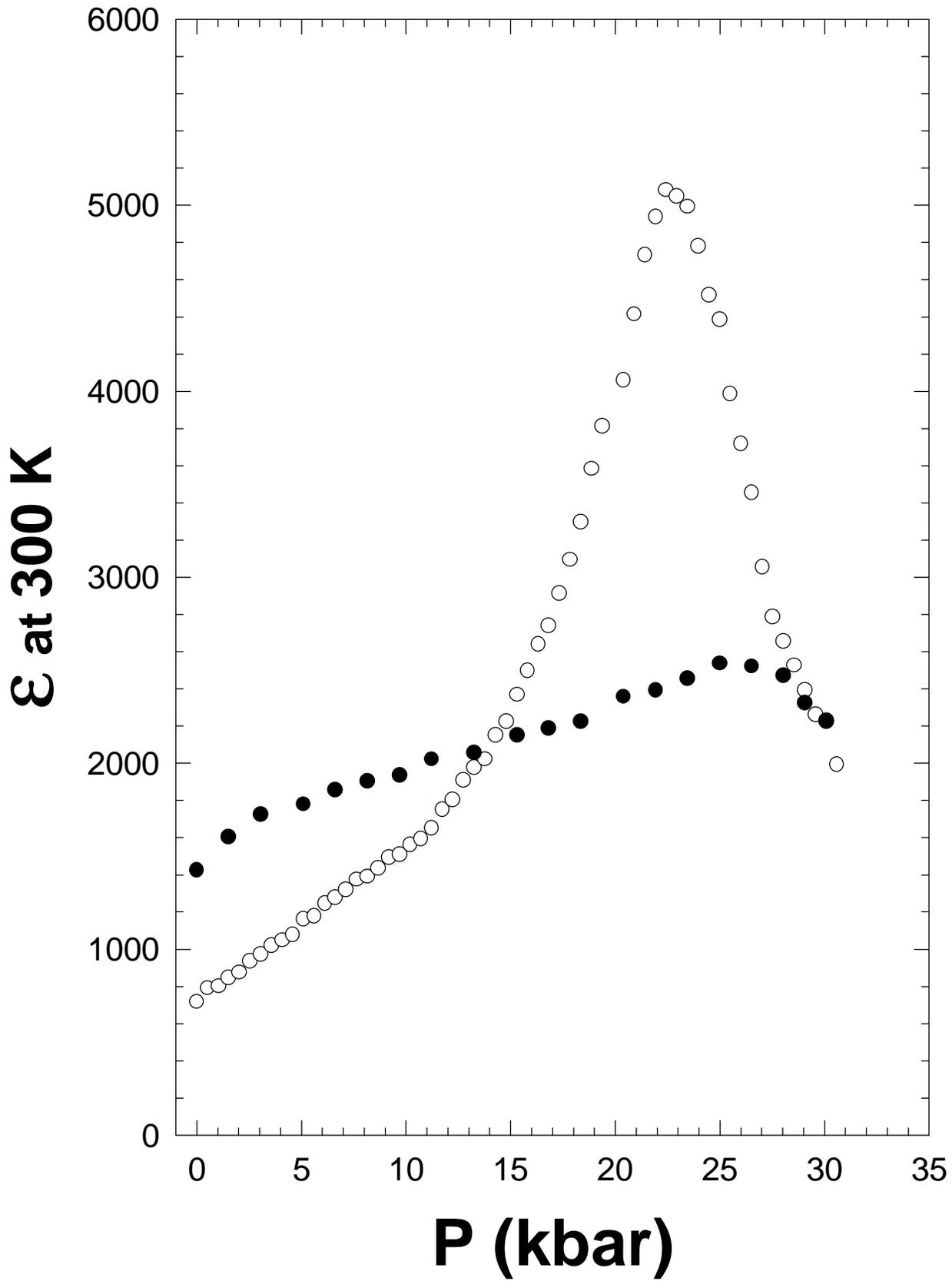
\end{document}